\documentclass[preprint,aps,draft]{revtex4}
\usepackage{epsf}
\begin{document}
\title{Comment on "Collective modes and gapped momentum states in liquid Ga: 
Experiment, theory, and simulation"
}

\author{Taras Bryk$^{1,2}$,
        Ihor Mryglod$^{1}$,
        Giancarlo Ruocco$^{3,4}$}

\affiliation{ $^1$ Institute for Condensed Matter Physics,National
Academy of Sciences of Ukraine,\\UA-79011 Lviv, Ukraine}
\affiliation{$^2$Institute of Applied Mathematics and Fundamental
Sciences,\\Lviv National Polytechnic University, UA-79013 Lviv,
Ukraine } 
\affiliation{$^3$
Center for Life Nano Science @Sapienza, Istituto Italiano di
  Tecnologia, 295 Viale Regina Elena, I-00161, Roma, Italy}
\affiliation{$^4$ Dipartimento di Fisica, Universita' di
Roma "La Sapienza", I-00185, Roma, Italy} 

\date{\today}
\begin{abstract}
We show that the presented in \cite{Khu20} theoretical expressions for longitudinal current 
spectral function $C^L(k,\omega)$ and dispersion of collective excitations are not correct.
Indeed, they are not compatible with the continuum limit and $C^L(k,\omega\to 0)$ contradicts the 
continuity equation.
\end{abstract}

\maketitle

\section{Introduction}

In a recent paper \cite{Khu20} the authors formulated their "overarching goal of this research 
programme ... to reach the stage where, despite the complexity of their theoretical description, 
liquids emerge as systems amenable to theoretical understanding at the level comparable to gases 
and solids". Looking at Figs.4 and 5 of \cite{Khu20} one can really make sure that the authors 
of \cite{Khu20} reached their ambitious goal in perfect agreement between the proposed theory and
computer simulations. In this paper the authors proposed theoretical expressions for the 
longitudinal current spectral function $C^L(k,\omega)$, with $k$ and $\omega$ being wave number and 
frequency, and for the dispersion of longitudinal collective excitations $\omega_c^L(k)$. 
Their expressions for $C^L(k,\omega)$ (Eq.18) and $\omega_c^L(k)$ (Eq.20), as one can judge 
from their Figs.4 and 5, recover with high
precision the molecular dynamics (MD) data in a wide range of wave numbers and temperatures.  
The $C^L(k,\omega)$ in their theoretical scheme was obtained from a simple continued fraction 
shown in their Eq.11. Although the standard approach for description of collective dynamics in 
liquids is to represent the Laplace-transformed density-density time correlation function
as a continued fraction\cite{Cop75,Sco05}, in \cite{Khu20} the authors derived the continued 
fraction for 
the longitudinal currrent-current correlations. Applying different closures for the chain of 
memory functions like in \cite{Han,Sco00,Ome98} one can obtain formal solution for 
$C^L(k,\omega)$ within a precision of several its frequency moments. 

However, such an approach of Ref.\cite{Khu20} is not really consistent with the 
hydrodynamics\cite{Han,Boo}, which is a 
collection of local conservation laws. Any liquid system on the spatial scales much 
larger than the mean interatomic distance must behave similarly from the point of view of slow 
collective modes derived by fluctuations of conserved quantities. In \cite{Khu20} the proposed 
theoretical approch is developed
from a single conserved dynamic variable, longitudinal component of total momentum $J^L(k,t)$, 
which is the slowest dynamic variable in the presented approach. It is well known from the textbooks
\cite{Han,Boo} as well as from other multivariable approaches \cite{deS88,Mry95,Bry01} which
dynamic variables are responsible for description of the viscoelastic transition in dispersion 
of collective excitations \cite{Bry10,Bry11}. The theoretical approach \cite{Khu20} does not 
contain coupling of longitudinal current fluctuations with the fluctuations of other conserved 
quantities, namely density $n(k,t)$ and energy $e(k,t)$ ones. The energy (or heat) density 
fluctuations reflect specific for liquids fluctuations of local temperature\cite{Bry12}, and 
long-wavelength heat relaxation processes are responsible for the central Rayleigh peak of the 
dynamic structure
factor $S(k,\omega)$ for one-component liquids at sufficiently small wave numbers $k$. Outside 
the hydrodynamic regime the short-wavelength density fluctuations $n(k,t)$ reflect the processes
connected with structural relaxation and instead of heat relaxation form the leading contribution
to the central peak of $S(k,\omega)$ \cite{Bry01b,Bry01c,Bry11}.
The presence of heat and density relaxation, therefore, are essential ingredients for a correct 
description of the spectra, including the
propagating density fluctuations regions, which are the main target of Ref.\cite{Khu20}.

The poor theoretical approach presented in Ref.\cite{Khu20}, missing the coupling with the most
important for liquids slow processes, is an oversimplified theory. It is, therefore, difficult to
understand why it is able to reproduce to a very good degree of accuracy the molecular
dynamics (MD) data for 
$C^L(k,\omega)$ in some region of wave numbers as it is shown in their Fig.4 \cite{Khu20}. Moreover,
we were motivated to understand why their expressions were able to recover the adiabatic speed of 
sound in the long-wavelength region of their Fig.5. Our question was: is it possible within 
the proposed fit-free theoretical scheme to obtain in the long-wavelength limit the
propagating modes with adiabatic speed of sound $c_s$? The multivariable approaches based on the 
set of dynamic variables $\{J^L(k,t),\dot{J}^L(k,t),....\}$ usually can produce in the 
long-wavelength limit the propagating modes only in elastic regime with propagation speed being the 
high-frequency one
$c_{\infty}$ slightly renormalized due to the coupling to faster kinetic modes. No viscoelastic 
effects like positive sound dispersion can be expected in this theory.

Motivated by the surprisingly good agreement shown in their Fig.4 we will check the expressions 
(Eqs.17-20) of \cite{Khu20} and behavior of their "relaxation parameters"
$\Delta_i(k)$ in the $k\to 0$ limit using a simple Lennard-Jones fluid, because of its simplicity 
in order to have analytical spacial derivatives of inteparticle potential needed for calculations of 
$\Delta_i(k)$ and their Eqs.17-20. In the next Section we provide details of our MD simulations 
and calculations of corresponding correlators. Then we will present our resuts and discuss them 
in comparison with the  Eqs.17-20 of \cite{Khu20}. The last Section contains conclusion of this 
study.

\section{Details of MD simulations}

We performed molecular dynamics simulations for supercritical Ne 
at T=295~K and density 1600 kg/m$^3$ using its Lennard-Jones potentials 
the same as in our previous study \cite{Bry17}. A model system of 4000 particles was simulated in 
microcanonical ensemble with perfect energy conservation over the whole production run of 300 000
time steps. The time step was 0.5 fs. Our main task was in sampling the space-Fourier components of 
all hydrodynamic variables, i.e. of density $n(k,t)$, mass-current ${\bf J}(k,t)$ and energy $e(k,t)$,
as well as of their time derivatives, in particular, of the mass-current up to the third order 
$\stackrel{\ldots}{\bf J}(k,t)$. We sampled all the possible wave vectors corresponding to the 
same absolute value, and used all them in spherical average of the corresponding correlators.
The smallest wave number sampled in this MD study was 0.143598\AA$^{-1}$.

In order to check reliability of the sampled time derivatives of the longitudinal mass-current 
and of our calculated static correlators we made use of the exact 
relations, which follow from a property of time derivatives of time correlations functions \cite{Han}
$$
\langle \dot{J}_L(-k)\dot{J}_L(k)\rangle\equiv -\langle \ddot{J}_L(-k) J_L(k)\rangle
$$
$$
\langle \ddot{J}_L(-k)\ddot{J}_L(k)\rangle\equiv -\langle \stackrel{\ldots}{J}_L(-k)\dot{J}_L(k)\rangle~.
$$
One can see in Fig.1 that perfect equivalence (difference less than 0.2\% for any $k$-point) is the
evidence of correct direct sampling of $J_L(k,t)$, $\dot{J}_L(k,t)$ $\ddot{J}_L(k,t)$ and 
$\stackrel{\ldots}{J}_L(k,t)$ in MD simulations. These dynamic variables are needed for calculations
of quantities $\Delta_i(k),~i=1,2,3$ in expressions for $C^L(k,\omega)$ and $\omega_c^L(k)$ in 
\cite{Khu20}.
Throughout this paper we will use reduced units of energy $k_BT=1$, mass $m=1$ and time 
$\tau_{\sigma}=1.997446 ps$
\begin{figure}\label{Fig1}
\epsfxsize=.40\textwidth {\epsffile{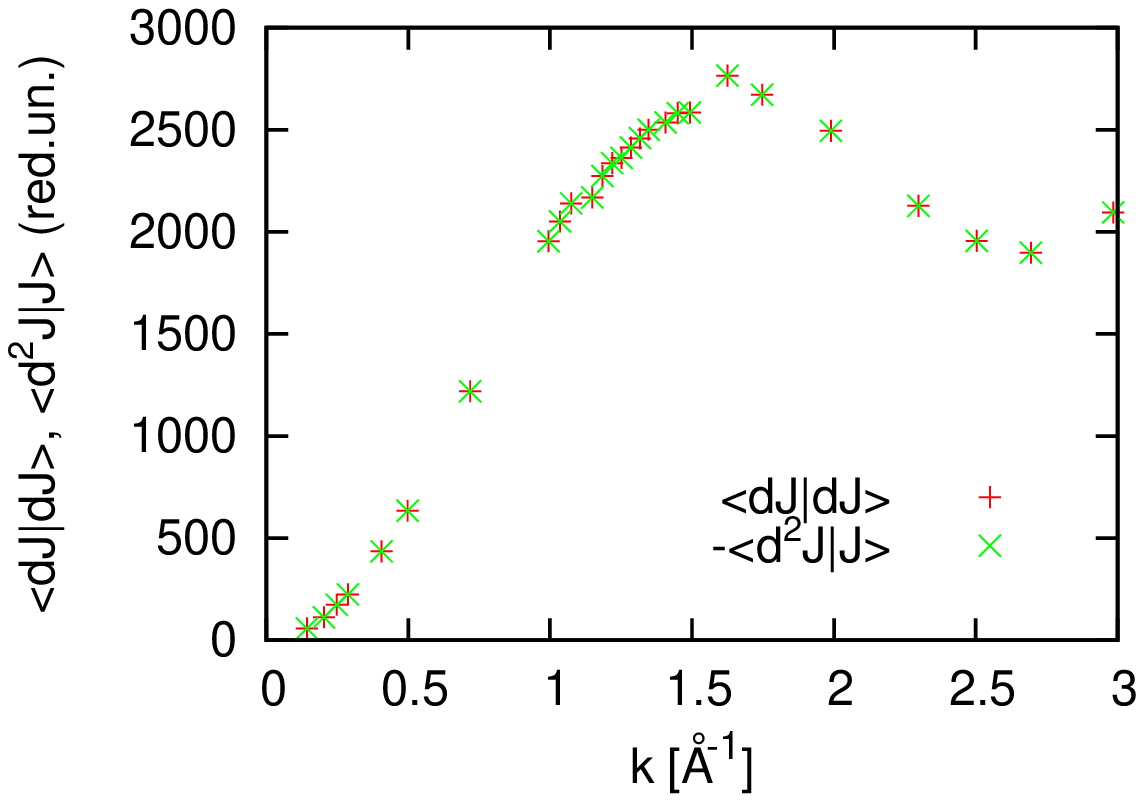}}
\epsfxsize=.40\textwidth {\epsffile{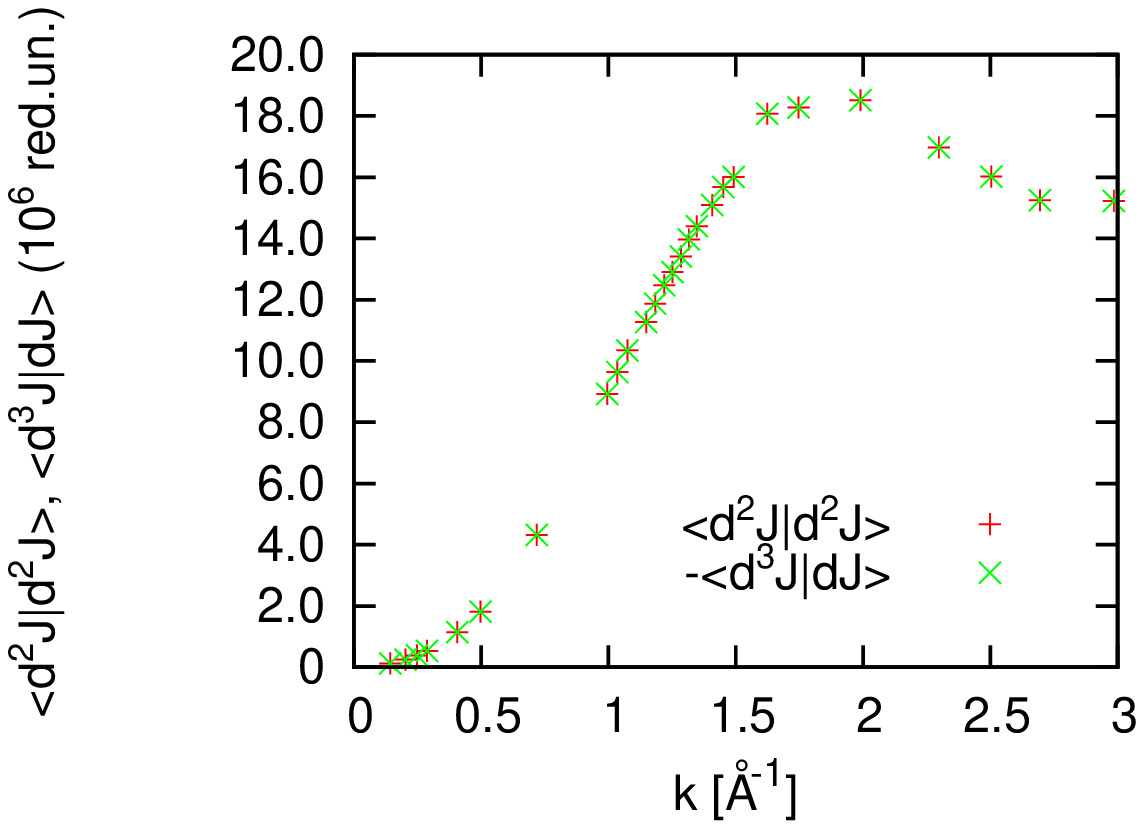}}
\caption{Check of the properties of time derivatives of longitudinal current for 
static correlators $\langle \dot{J}_L(-k)\dot{J}_L(k)\rangle\equiv -\langle \ddot{J}_L(-k)
J_L(k)\rangle$ (a) and $\langle \ddot{J}_L(-k)\ddot{J}_L(k)\rangle\equiv -\langle 
\stackrel{\ldots}{J}_L(-k)
\dot{J}_L(k)\rangle$ (b) for supercritical Ne at T=295~K and density 1600 kg/m$^3$.}
\end{figure}

\section{Results and discussion}

As we mentioned above the perfect agreement between the proposed in \cite{Khu20} fit-free theory 
and MD results for $C^L(k,\omega)$ in their Fig.4 looks too good to be true. Indeed, a simplest 
check of their Eq.18 in the $\omega\to 0$ limit results in the non-zero value of $C^L(k,\omega=0)$
\begin{equation}
C^L(k,\omega=0)=\frac{1}{\pi}\frac{\Delta_1(k)\Delta_2(k)\Delta_3(k)^{3/2}}{B_0(k)}
\equiv \frac{1}{\pi}\frac{\Delta_2(k)}{\Delta_1(k)\Delta_3(k)^{1/2}}~,
\end{equation}
while any viscoelastic theory must result in $C^L(k,\omega=0)\equiv 0$ as the consequence of 
continuity equation. We cannot explain how the authors \cite{Khu20} obtained in their Fig.4 
the $C^L(k,\omega\to 0)\propto \omega^2$ behavior from their fit-free theory (their Eq.18). 

We calculated from their Eqs.16-17  the "relaxation parameters" $\Delta_i(k),~i=1,2,3$ and 
doublechecked the relations:
$$
\Delta_1(k)+\Delta_2(k)=\frac{\langle \ddot{J}(-k)\ddot{J}(k)\rangle}
                             {\langle \dot{J}(-k)\dot{J}(k)\rangle}~,
$$
where the right hand side tends to a constant in long-wavelength limit and is simply the ratio 
of $k$-dependences shown in Fig.1(a,b),
and
$$
\Delta_3(k)=[\langle\stackrel{\ldots}{J}(-k)\stackrel{\ldots}{J}(k)\rangle 
  -\frac{\langle \ddot{J}_L(-k)\ddot{J}_L(k)\rangle^2}{\langle \dot{J}_L(-k)\dot{J}_L(k)\rangle}]
  /[\langle \ddot{J}_L(-k)\ddot{J}_L(k)\rangle 
  -\frac{\langle \dot{J}_L(-k)\dot{J}_L(k)\rangle^2}{\langle J_L(-k)J_L(k)\rangle}].
$$
In Fig.2 we show the $k$-dependence of the "relaxation parameters"\cite{Khu20} and one can 
see the parameters $\Delta_2(k)$ and $\Delta_3(k)$ tending in the long-wavelength limit to non-zero 
values while 
$$
\Delta_1(k)\equiv \frac{\langle \dot{J}_L(-k)\dot{J}_L(k)\rangle}{\langle J_L(-k)J_L(k)\rangle}
$$ 
behaves in $k\to 0$ limit as $\propto c_{\infty}^2k^2$ with $c_{\infty}$
being the high-frequency speed of sound.
\begin{figure}\label{Fig2}
\epsfxsize=.50\textwidth {\epsffile{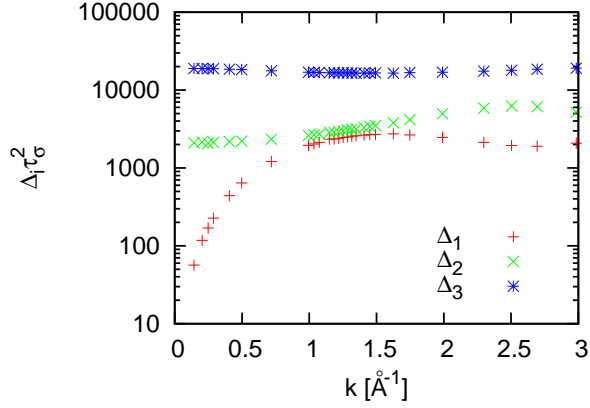}}
\caption{Dependence of the "relaxation parameters" $\Delta_i,~i=1,2,3$ (Eq.16 of \cite{Khu20}) 
on wave numbers for supercritical Ne at T=295~K and density 1600 kg/m$^3$. }
\end{figure}

Now we can estimate how large is the deviation of $C^L(k,\omega=0)$ from the correct zero value.
Since the $\Delta_1(k)$ goes to zero in the long-wavelength limit and $\Delta_2(k\to 0)$ 
and $\Delta_3(k\to 0)$ tend to finite non-zero values, the resulting $C^L(k,\omega=0)$ taken from 
Eq.18 of \cite{Khu20} should diverge for $k\to 0$. Indeed, in Fig.3 one can observe the strong 
increase of $C^L(k,\omega=0)\propto k^{-2}$ in \cite{Khu20}, that means wrong theoretical result 
comparing with the exact relation $C^L(k,\omega\to 0)=0$.
\begin{figure}\label{Fig3}
\epsfxsize=.50\textwidth {\epsffile{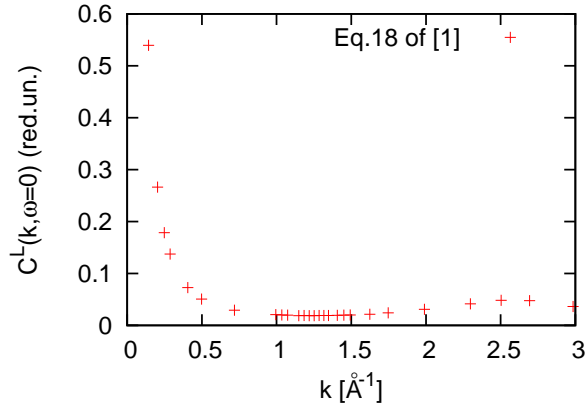}}
\caption{Dependence of the zero-frequency value $C^L(k,\omega=0)$ (Eq.18 of \cite{Khu20}) 
on wave numbers for supercritical Ne at T=295~K and density 1600 kg/m$^3$.}
\end{figure}

Now we will analyze the expression for dispersion of collective excitations \cite{Khu20}.
Since only the "relaxation parameter" $\Delta_1(k)$ tends to zero as $k^2$ in the long-wavelength
limit, and higher "relaxation parameters"  $\Delta_{2,3}(k)$ tend to constants in that limit,
one can easily estimate, that their Eq.(20) for $\omega_c^L(k)$ tends to a constant for $k\to 0$
$$
\omega_c^L(k\to 0)=\frac{\Delta_2(0)}{\sqrt{2[\Delta_3(0)-\Delta_2(0)]}}~,
$$
while the correct dispersion law had to recover in that limit the hydrodynamic dispersion law 
$\omega(k\to 0)=c_sk$.
In Fig.4 we show the dispersion of collective acoustic modes estimated from the peak 
positions of MD-derived $C^L(k,\omega)$ (plus symbols with error bars) and compare it with 
the dispersion of "bare" (non-damped) high-frequency modes which in the long-wavelength 
limit have linear dispersion with the high-frequency (elastic) speed of sound $c_{\infty}$
\begin{equation}\label{high-fr}
\omega_{\infty}(k\to 0)=[\frac{\langle\dot{J}_L(-k)\dot{J}_L(k)\rangle}
{\langle{J}_L(-k){J}_L(k)\rangle}]^{1/2}|_{k\to 0} \to c_{\infty}k~.
\end{equation}
The coupling to the faster dynamic modes (connected with higher time derivatives of the longitudinal
current) can only slightly renormalize down the theoretical dispersion law, however it will 
never result in the hydrodynamic speed of sound $c_s$ and positive sound dispersion \cite{Bry10}.
Within the proposed in \cite{Khu20} theoretical approach is impossible to obtain the propagating 
modes with adiabatic speed of sound, because in order to obtain it one has to include coupling with 
density and energy (or heat) density fluctuations into the theoretical scheme. 
And, as it was expected from the wrong 
behavior of $C^L(k,\omega)$ discussed above, the proposed expression for dispersion of longitudinal
collective excitations is wrong too. In Fig.4 only for two lowest $k$-values we obtained the 
positive expression under the square root in their Eq.(20). For higher wave numbers the expression
under the square root became negative, i.e. no propagating modes for those wave numbers. It is not 
clear how in Fig.5 of \cite{Khu20} the authors were able to reproduce perfectly the MD data 
by using their Eq.20 and even reach the adiabatic speed of sound in the long-wavelength region, 
that is impossible to do in their theoretical approach. Even conceptually their theoretical approach,
 which
does not contain coupling to fluctuations of conserved quantities, density $n(k,t)$ and energy 
density $e(k,t)$, and Eq.20 cannot result in the 
long-wavelength limit in the linear dispersion with the adiabatic speed of sound.
In their run for the "overarching goal of this research programme" the authors forgot about the 
existing methodologies of calculations and theories of collective excitations in liquids, which 
correctly satisfy exact relations and a large number of sum rules.
\begin{figure}\label{Fig4}
\epsfxsize=.50\textwidth {\epsffile{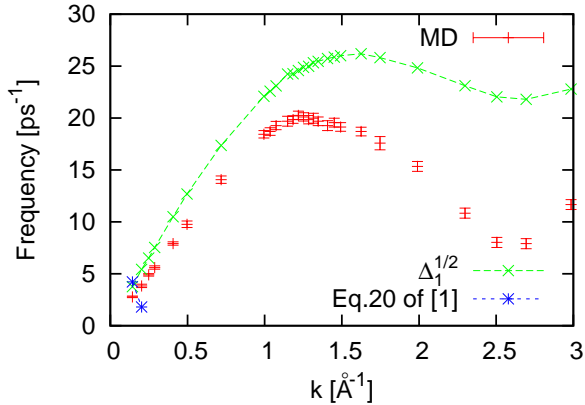}}
\caption{Peak positions of the longitudinal current spectral function
$C^L(k,\omega)$, obtained from MD simulation (plus symbols with error bars). 
The dispersion of the nondamped high-frequency acoustic-like modes with 
long-wavelength asymptote (\ref{high-fr}) is shown by line-connected 
cross symbols.
Eq.20 of \cite{Khu20} (line-connected star symbols) contains positive 
expression unders square root only for two lowest $k$-points, for larger 
$k$-values no real $\omega_c^{L}(k)$ exist. 
}
\end{figure}

Another point we want to discuss here is the claimed "gapped momentum states"\cite{Khu20}. 
It sounds strange that the authors are trying to represent the well known in the literature 
shear waves with a propagation gap as some special finding and rename them as the 
"gapped momentum states". The title of \cite{Khu20} stating "Collective modes and 
gapped momentum states ..." clearly discriminates between the "collective modes" and 
"gapped momentum states" that is not correct because there is no difference between 
ordinary collective 
shear waves and "gapped momentum states".  Moreover, it has been known for long time that other 
collective propagating processes in liquids have very similar behavior of their dispersion, like 
heat waves \cite{Jos89,Bry01,Bry11} or optic-like modes in binary liquids with demixing tendencies 
\cite{Bry02,Bry11}.  We would like
to remind the readers that by 2017 the same group assured the community in Frenkel-like dispersion
of the transverse excitations in liquids\cite{Bra12,Wan17}, i.e. when the transverse excitations 
in liquids 
exist only above the so-called Frenkel frequency cut-off, that contradicted the existed theories of 
transverse exsitations\cite{Han,Mac84,Bry00} and MD data (see our discussion in \cite{Bry17}), which
evidenced on existing the long-wavelength propagation gap for shear waves.
In 2017 the same authors revealed that the dispersion of shear waves indeed starts from zero 
frequency outside the propagation gap and published a paper \cite{Yan17} in which claimed that the 
propagation gap originates from the Frenkel jumps and is defined by the 
single-particle Frenkel time (quoting \cite{Yan17}: "$\tau$ is understood to be the full period of 
the particles’ jump motion equal to twice Frenkel’s $\tau$"). 
That claim again contradicted the existed theory of 
transverse excitations in liquids\cite{Han,Mac84,Bry00}, in which the collective shear stress 
relaxation with Maxwell relaxation time is responsible for the propagation gap and we showed 
several times that there is huge difference between the collective and single-particle relaxation
processes in their effect on transverse dynamics\cite{Bry18,Bry18b}. Now, in \cite{Khu20} the same 
group started to rename the ordinary shear waves of liquid dynamics as "gapped momentum states".

\section{Conclusion}

The proposed in \cite{Khu20} theoretical scheme for description of longitudinal collective 
excitations in simple liquids is not consistent with hydrodynamics, because only one hydrodynamic
variable, the longitudinal  current, was used in that scheme, that rised questions whether the 
obtained in \cite{Khu20} expressions for longitudinal current spectral function $C^L(k,\omega)$
and for the dispersion of collective excitations are correct.
We performed molecular dynamics simulatins on a simple supercritical Ne at 295K and density 
1600 kg/m$^3$ with a purpose of numerical check of these expressions. 

We showed that the proposed in \cite{Khu20} expression for $C^L(k,\omega)$ does not have 
correct low-frequency limit $C^L(k,\omega\to 0)$ and even diverges in the long-wavelength limit,
that is wrong, while according to the continuity equation it must be $C^L(k,\omega=0)\equiv 0$.
Why their Fig.4 shows perfect agreement of their theoretical $C^L(k,\omega)$ with MD data we 
cannot explain.

Within the proposed in \cite{Khu20} theoretical scheme it is impossible to recover the 
hydrodynamic dispersion law in $k\to 0$ limit and macroscopic
adiabatic speed of sound, because the coupling of the longitudinal current with other fluctuations
of conserved quantities is absent in that scheme. We checked the proposed in \cite{Khu20} 
expression for the dispersion of collective excitations and found that with increasing wave 
numbers the expression under square root in their Eq.20 becomes negative, i.e. wrong result.
Why their Fig.5 shows perfect agreement between their theoretical expression and the MD-obtained
dispersion of collective excitation, and even recovers the hydrodynamic linear dispersion law 
with $c_s$ we cannot explain.
We would suggest the authors of \cite{Khu20} to show their similar checks for the correlators
$\langle \ddot{J}_L(-k) J_L(k)\rangle$ and $\langle \stackrel{\ldots}{J}_L(-k)\dot{J}_L(k)\rangle$
as we presented in Fig.1, as well as to reveal the $k$-dependence of their $\Delta_i(k)$. This 
defintely will allow to find out why the low-frequency limit of $C^L(k,\omega)$, their Eq.18, and 
the long-wavelength limit of $\omega^L_c(k)$, their Eq.20, do not correspond to the data in their
Figs.4 and 5, respectively.

\end{document}